\documentclass[aps,prl,reprint,floatfix,longbibliography,superscriptaddress]{revtex4-2}
\usepackage{graphicx,float}
\usepackage{amsmath,amssymb,mathrsfs,dcolumn}
\usepackage{gensymb}
\usepackage[usenames]{color}
\usepackage{subfigure}
\usepackage[normalem]{ulem} 
\usepackage{hyperref}
\usepackage{verbatim}
\hypersetup{colorlinks=true, linkcolor=blue, citecolor=red, urlcolor=blue}

\begin{document}
\title{Pure Dephasing of Magnonic Quantum States}
\author{H. Y. Yuan}
\author{W. P. Sterk}
\affiliation{Institute for Theoretical Physics, Utrecht University, 3584CC Utrecht, The Netherlands}
\author{Akashdeep Kamra}
\affiliation{Condensed Matter Physics  Center (IFIMAC) and Departamento de F\'{i}sica Te\'{o}rica de la Materia Condensada, Universidad Aut\'{o}noma de Madrid, E-28049 Madrid, Spain}
\author{Rembert A. Duine}
\affiliation{Institute for Theoretical Physics, Utrecht University, 3584CC Utrecht, The Netherlands}
\affiliation{Department of Applied Physics, Eindhoven University of Technology, P.O. Box 513, 5600 MB Eindhoven, The Netherlands}
\date{\today}

\begin{abstract}
For a wide range of nonclassical magnonic states that have been proposed and demonstrated recently, a new time scale besides the magnon lifetime -- the magnon dephasing time -- becomes important, but this time scale is rarely studied. Considering exchange interaction and spin-phonon coupling, we evaluate the pure magnon dephasing time and find it to be smaller than the magnon lifetime at temperatures of a few Kelvins. By examining a magnonic cat state as an example, we show how pure dephasing of magnons destroys and limits the survival of quantum superpositions. Thus it will be critical to perform quantum operations within the pure dephasing time. We further derive the master equation for the density matrix describing such magnonic quantum states taking into account the role of pure dephasing, whose methodology can be generalized to include additional dephasing channels that experiments are likely to encounter in the future. Our findings enable one to design and manipulate robust quantum states of magnons for information processing.
\end{abstract}

\maketitle


{\it Introduction.---} Rapid advancements in control and manipulation of magnons -- spin excitations of ordered magnets -- have witnessed the experimental generation of their nonclassical states~\cite{Lachance-QuirionSciAdv2017,Lachance-QuirionScience2020,Lachance-QuirionAPE2019,YuanArxiv2021}. Strong magnonic coupling to superconducting qubits~\cite{TabuchiScience2015,Lachance-QuirionAPE2019}, photons~\cite{HueblPRL2013,ZhangPRL2014,HarderSSP2018,RameshtiArxiv2021}, and phonons~\cite{WeilerPRL2011,RuckriegelPRB2014,HolandaNatPhys2018,BozhkoLTP2020} has been demonstrated. This has triggered a wide range of theoretical proposals suggesting the use of magnonic nonclassical states in providing a resource of entanglement and memories for quantum communication and information processing~\cite{YuanArxiv2021,Agarwal2019,LiPRL2018,KamraPRB2019,YuanPRB2020B,ElyasiPRB2020,ZouPRB2020,SunPRL2021,PottsPRAppl2020,WuhrerArxiv2021,MousolouPRB2021}.

\begin{figure}[tbh]
\begin{center}
\includegraphics[width=85mm]{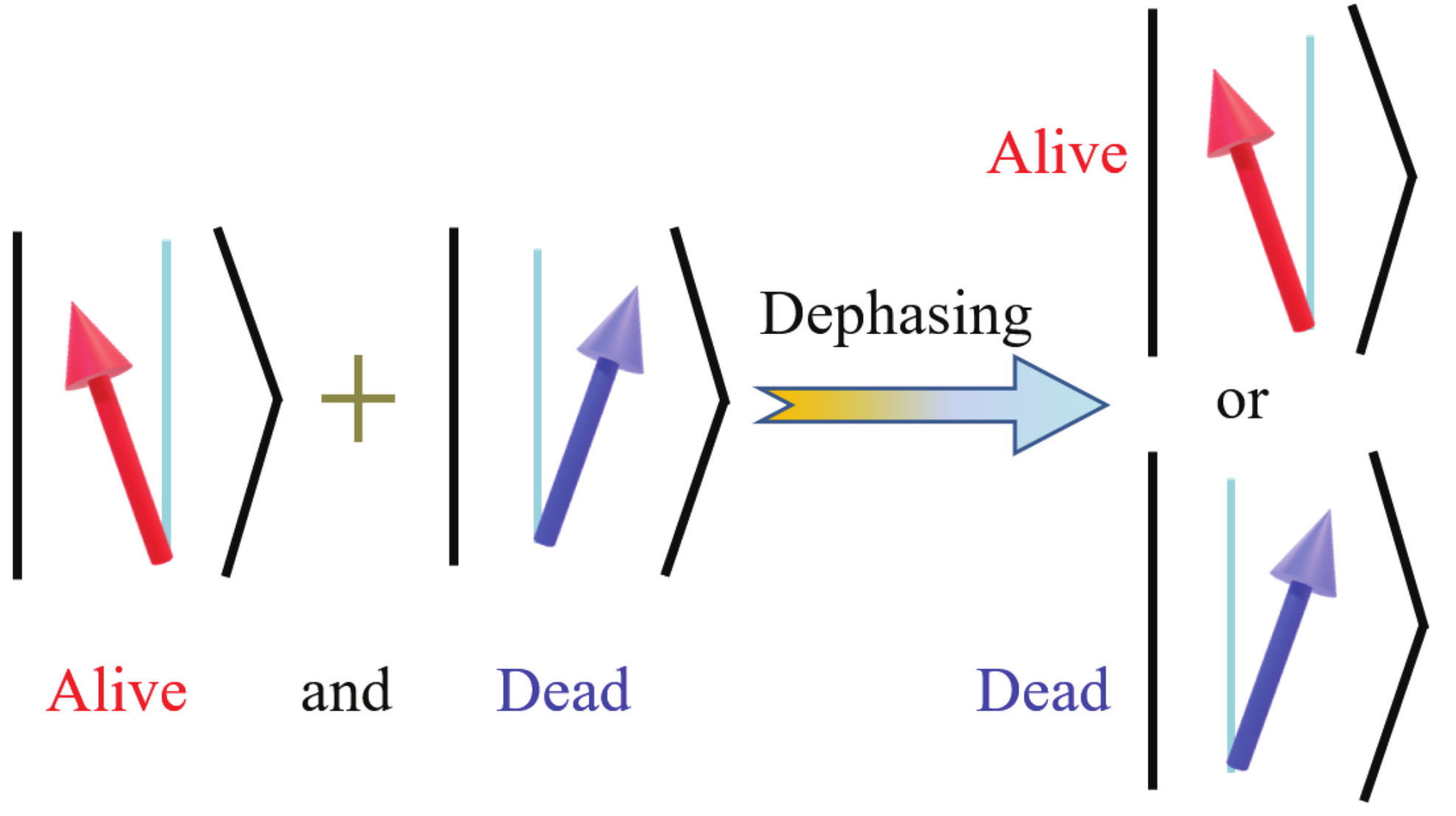}
\caption{Schematic depiction of a magnonic cat state and loss of its quantum superposition due to dephasing. The cat state has a left-tilting magnetic moment state (alive cat) superposed with its right-tilting state (dead cat). The magnetic moment thus points in two directions at the same time, a possibility not allowed in classical magnetization dynamics. Dephasing results in the extinguishing of such a quantum superposition and brings the system to one of its classically permitted states, with the magnetic moment pointing along a unique direction.}
\label{fig1}
\end{center}
\end{figure}

One of the key motivators driving this young field of quantum magnonics is the low magnon relaxation rates in several magnetic insulators. However, in considering nonclassical states of magnons, such as single-magnon~\cite{Lachance-QuirionScience2020,YuanPRB2020}, squeezed~\cite{ZhaoPRL2004,KamraPRL2016,LiPRA2019,KamraAPL2020}, and cat states~\cite{SharmaPRB2021,SunPRL2021} (Fig.~\ref{fig1}), a new time scale emerges -- the magnon dephasing time. This is because nonclassical states are comprised by a nontrivial quantum superposition of the various magnon number states~\cite{GerryBook2004}. This superposition is expected to be destroyed by dephasing (Fig.~\ref{fig1})~\cite{ZurekRMP2003,XiongPLA2019}. This time scale is the magnon equivalence of the well-known $T_2^\ast$ for qubits. Considering, for example, that $T_2^\ast$ of semiconductor-based spin qubits~\cite{VandersypenPT2019,ChatterjeeNatRevPhys2021,BurkardArxiv2021} is orders of magnitude smaller than their relaxation time $T_1$, we anticipate that the magnon dephasing time will become the limiting factor in the generation and usefulness of magnonic quantum states. Understanding magnon dephasing time is thus critical for progress in quantum magnonics~\cite{YuanArxiv2021}.

In the context of magnon spin transport~\cite{BauerNatMat2012,BrataasPR20201}, the role of various scattering processes that cause magnon relaxation has been investigated in great detail~\cite{KrivosikPRB2010,AkhiezerBook1968,CornelissenPRB2016}. The spin-conserving nature of the exchange interaction -- the strongest energy scale in the magnet -- reduces its relative importance for many of the transport and relaxation phenomena~\cite{CornelissenPRB2016,AkhiezerBook1968}. To understand dephasing, qualitatively different processes which leave the magnon in its original state become important~\cite{XiongPLA2019}. These channels do not affect magnon relaxation or transport, and thus, have been largely overlooked so far. A class of these dephasing phenomena is mediated by the strong exchange interaction, as we find here, giving them an important role.

In this Letter, we provide a derivation of the master equation for the density matrix describing the magnonic system duly taking into account pure dephasing. We focus on two microscopic mechanisms. First, the four-magnon processes mediated by exchange interaction are shown to underlie magnon dephasing caused by coupling all magnonic modes. The resulting dephasing rate is found to scale quadratically with temperature and exceeds the typical relaxation rate of magnetic insulators at around 1 K. Second, we consider the two-magnon-one-phonon processes due to spin-phonon coupling.
The dephasing rate due to this mechanism scales linearly with temperature and is comparable to the relaxation rate at a few Kelvins. Our methodology based on master equation can be extended to address other dephasing mechanisms in the near-future experiments.


\begin{figure}
	\centering
	\includegraphics[width=0.49\textwidth]{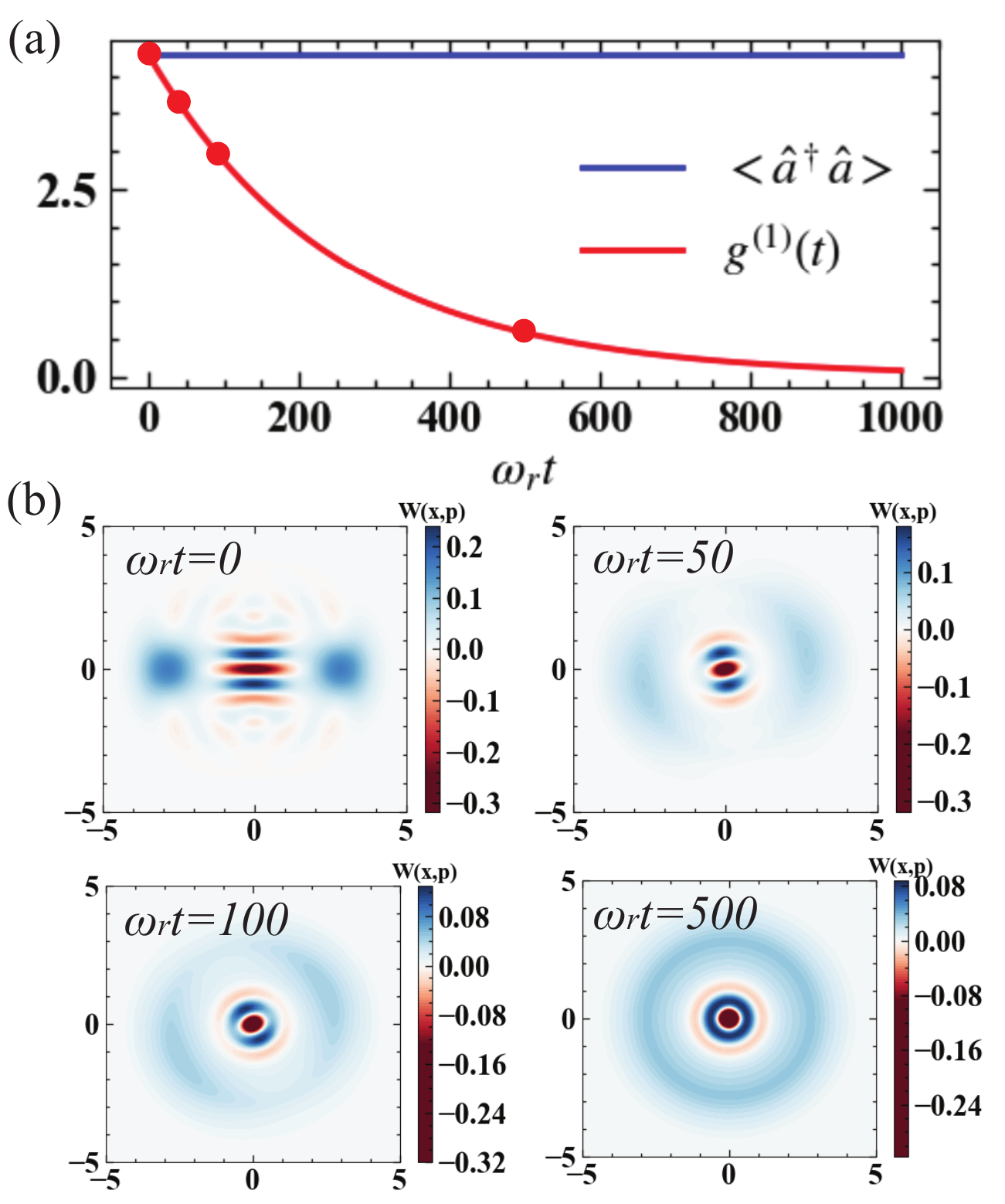}\\
	\caption{Pure dephasing of a magnonic cat state. (a) The coherence (red line) decreases with time while the magnon number (blue line) remains constant. This shows that pure dephasing does not cause relaxation of energy or magnon number, but it kills the quantum coherence of a state. (b) Wigner distributions of the magnonic state at $\omega_rt=0,50,100,500$, starting from an odd-parity cat state at $t = 0$. The two separable blue blobs visible at time $t = 0$ represent the macroscopically distinct alive and dead cat. At large times, the blobs have merged into one annulus and the quantum superposition of the cat state has been removed, resulting in a mixed state. The parameters are temperature $T=1~\mathrm{K}$, external field $H=1~\mathrm{T}$, pure dephasing rate $\gamma/\omega_r=0.002$, and amplitude of the initial cat state $\beta=2$.}\label{fig2}
\end{figure}

{\it Pure dephasing of a magnonic cat state.---} Before delving into the master equation derivation and the dephasing rates, as an example, we examine the effect of pure dephasing on a magnonic cat state~\cite{SharmaPRB2021,SunPRL2021,DodonovPhysica1974,SchlomannPR1961}, schematically depicted in Fig.~\ref{fig1}. This serves to bring out the quantum nature of the dephasing process clearly. Dephasing is expected to strongly influence the coherence of various quantum states of magnons, including single-magnon state, squeezed states, and cat states. Here, as an example, we examine how an odd-parity cat state $|\psi_0\rangle=(|\beta\rangle - |-\beta \rangle)/\sqrt{2}$ dephases, where $|\beta\rangle$ is a coherent state of magnons with amplitude $\beta$, defined as $\hat{a}|\beta\rangle = \beta |\beta\rangle$ with $\hat{a}$ being the annihilation operator of a particular magnon mode. To characterize the coherence of the state \cite{Scullybook1997}, we employ the unnormalized first-order coherence function $g^{(1)}(t)= \langle \hat{a}(t)\hat{a}(0)\rangle$. By numerically solving the master equation \eqref{master_equation}, derived below, we find it decays monotonically with time [see Fig.~\ref{fig2}(a)]. On account of the pure magnon dephasing considered here, the coherence of magnons (red line) is found to gradually decay with time without any relaxation of the magnon number (blue line). This decoherence process is also observed in phase space as shown in Fig.~\ref{fig2}(b). At time $t = 0$, the Wigner function of the magnonic cat state shows a clear two-blobs feature of quantum superposition between the two coherent states, each representing the cat in a different state. As the time goes by, the two distinct blobs merge into each other resulting in a thermal state devoid of the quantum superposition characteristic of the cat state.


{\it Dephasing due to magnon-magnon interaction.---} We now derive the master equation that describes the time evolution of the magnon density matrix, starting from the microscopic Hamiltonian for a Heisenberg ferromagnet. To this end, we consider
\begin{equation}\label{eq_Heisenberg_Ham}
\hat{\mathcal{H}}=-J \sum_{\langle i,j \rangle } \hat{\mathbf{S}}_i \cdot \hat{\mathbf{S}}_j - H \sum_i \hat{\mathbf{S}}_{iz},
\end{equation}
where $J$ parameterizes the exchange interaction between neighboring spins, $\hat{\mathbf{S}}_i$ is the spin operator on the $i$-th site with spin number $S$, and $H$ is an external magnetic field oriented along the $z$-axis. The ground state of the system is a ferromagnetic state $\mathbf{S}_i=S\mathbf{e}_z$. Employing the Holstein-Primakoff (HP) transformation~\cite{HolsteinPRB1940}, the magnon excitations are related to the spin raising and lowering operators as:  $S_i^+= \sqrt{2S-\hat{a}_i^\dagger \hat{a}_i}\hat{a}_i,~S_i^-=\hat{a}_i^\dagger \sqrt{2S-\hat{a}_i^\dagger \hat{a}_i},~S_{iz}=S-\hat{a}_i^\dagger \hat{a}_i$, where $\hat{a}_i$ ($\hat{a}^\dagger_i$) is the magnon annihilation (creation) operator on $i-$th site which satisfies the commutation relation $[\hat{a}_i, \hat{a}_j^\dagger ] = \delta_{ij}$, and $S_i^\pm \equiv S_{ix} \pm iS_{iy}$. Employing these relations, the Hamiltonian \eqref{eq_Heisenberg_Ham} is written in Fourier space up to the fourth order in the magnon ladder operators as
\begin{equation} \label{heisenberg_ham}
\hat{\mathcal{H}}=\sum_\mathbf{k} \omega_\mathbf{k} \hat{a}_\mathbf{k}^\dagger \hat{a}_\mathbf{k} + \sum_{\mathbf{k},\mathbf{k}',\mathbf{q}} C(\mathbf{k},\mathbf{k}',\mathbf{q}) \hat{a}_\mathbf{k+q}^\dagger \hat{a}_\mathbf{k'-q}^\dagger \hat{a}_\mathbf{k'} \hat{a}_\mathbf{k},
\end{equation}
where $\omega_\mathbf{k}=2JSd^2\mathbf{k}^2+H$ is the magnon dispersion $d$ the lattice constant, and $C(\mathbf{k},\mathbf{k}',\mathbf{q})$ is the scattering amplitude that is proportional to exchange coefficient $J$ and depends on the structure factor of lattice. Focusing on the dynamics of a specific mode with wavevector $\mathbf{k}_0$, we can simplify the four-magnon term in \eqref{heisenberg_ham} and rewrite the Hamiltonian as
\begin{equation} \label{exchange_ham}
\hat{\mathcal{H}}=\omega_r \hat{a}^\dagger \hat{a} +\sum_{\mathbf{k} \neq \mathbf{k}_0} \omega_\mathbf{k} \hat{a}_\mathbf{k}^\dagger \hat{a}_\mathbf{k} + \hat{a}^\dagger \hat{a} \sum_{\mathbf{k} \neq \mathbf{k}_0} g(\mathbf{k})\hat{a}_\mathbf{k}^\dagger \hat{a}_\mathbf{k},
\end{equation}
where we have dropped the subscript $\mathbf{k}_0$ of $\hat{a}$ to keep the notation simple, $\omega_r \equiv \omega_{\mathbf{k}_0}$ and $g(\mathbf{k}) \equiv C(\mathbf{k}_0,\mathbf{k},\mathbf{q}=0)$ is the coupling strength between mode $\mathbf{k}_0$ and other magnon modes. To consider how the scattering of the $\mathbf{k}_0$ mode with other magnon modes influences the dynamics of the $\mathbf{k}_0$ mode, we treat all the other magnons as a bath~\cite{XiongPLA2019}.

The total magnonic system is Hermitian and the density matrix thus satisfies the Schr\"{o}dinger equation in the interaction picture as $d\tilde{\rho}_T/dt = -i[\tilde{\mathcal{H}}_{\mathrm{int}},\tilde{\rho}_T]$, where $\tilde{\rho}_T$ is the total density matrix of the system, $\tilde{\mathcal{H}}_{\mathrm{int}}=\hat{a}^\dagger \hat{a} \sum_{\mathbf{k} \neq \mathbf{k}_0} g(\mathbf{k})\hat{a}_\mathbf{k}^\dagger \hat{a}_\mathbf{k}$ is the interacting Hamiltonian, and the tilde decorates an operator in the interaction picture. The density matrix of the $\mathbf{k}_0$ mode is obtained by tracing all the other degrees of freedom, i.e. $\tilde{\rho}=tr_R(\tilde{\rho}_T)$. Within the Born-Markov approximation, the evolution of $\tilde{\rho}$ is written as~\cite{ManzanoAIP2020}
\begin{equation}
\frac{d\tilde{\rho}}{dt}=-\int_0^t dt' tr_R \left ( [\tilde{\mathcal{H}}_{\mathrm{int}}(t'),[\tilde{\mathcal{H}}_{\mathrm{int}}(t'),\tilde{\rho}(t) \otimes \tilde{R}_0]]\right ),
\end{equation}
where $\tilde{R}_0$ refers to the initial density matrix of the bath. By substituting \eqref{exchange_ham} into the master equation above, after a tedious but straightforward calculation, we obtain the dynamical equations for the magnon mode of interest in the Schr\"{o}dinger picture as
\begin{equation}\label{master_equation}
\frac{d \hat{\rho}}{dt}=i[\hat{\rho}, \hat{\mathcal{H}}_s] + \gamma \mathcal{L}_{\hat{n}\hat{n}}(\hat{\rho}),
\end{equation}
where $\hat{\rho}$ is the density matrix describing the mode $\mathbf{k}_0$, $\hat{\mathcal{H}}_s=\omega_r \hat{a}^\dagger \hat{a}$, the Liouville superoperator $\mathcal{L}_{\hat{n}\hat{n}}(\hat{\rho}) \equiv 2\hat{n} \hat{\rho} \hat{n}^\dagger - \hat{n}^\dagger \hat{n} \hat{\rho} - \hat{\rho} \hat{n}^\dagger \hat{n} $ with $\hat{n}=\hat{a}^\dagger \hat{a}$. The parameter $\gamma$ characterizes the decoherence/dephasing rate and is given by
\begin{equation}\label{exchange_integral}
\gamma = \int_0^\infty d\omega D^2(\omega) |g(\omega)|^2 n_{\mathrm{th}}[n_{\mathrm{th}}+1],
\end{equation}
where $n_{\mathrm{th}} = [\exp(\omega/k_BT)-1]^{-1}$ is the Bose-Einstein distribution, $k_B$ is Boltzmann constant, and $D(\omega)$ is the magnon density of states.

Employing the master equation \eqref{master_equation}, we immediately see that the average magnon number $\langle \hat{n} \rangle = tr(\hat{\rho} \hat{n})$ does not change with time since it commutes with the Hamiltonian $\hat{\mathcal{H}}_s$, i.e., there is no energy relaxation. On the other hand, employing Eq.~\eqref{master_equation}, the first moment $\langle \hat{a} \rangle$ evolves as
\begin{equation}
\frac{d\langle \hat{a} \rangle}{dt}= -i (\omega_r-i\gamma) \langle \hat{a} \rangle.
\end{equation}
The solution of this equation is $\langle \hat{a}(t) \rangle = \langle \hat{a}(0) \rangle \exp{(-i\omega_rt -\gamma t)}$, which implies that the coherence of the system $g^{(1)}(t)$ will be lost with a time scale of $T_2^*=1/\gamma$ [see Fig.~\ref{fig2}(a)]. Since this process is not accompanied by magnon number relaxation [Fig.~\ref{fig2}(a)], we call it a pure magnon dephasing process.

An intuitive understanding of this magnon dephasing process is achieved by contrasting it with pure dephasing in a qubit~\cite{NazarovBook2009}. According to Eq.~\eqref{exchange_ham}, the scattering of mode $\mathbf{k}_0$ with other magnons adds a random fluctuation $\zeta$ to the eigenfrequency $\omega_r$, while the amplitude of the magnon mode is not changed. After a sufficiently long time, even though the average frequency is still $\omega_r$, the phase fluctuations of the magnon mode will vary with time as $\delta \varphi \propto \sqrt{t}$. This is similar to the random walk of a Brownian particle \cite{FeynmanBook2015}. When the phase uncertainty $\delta \varphi$ exceeds $2\pi$, the magnon mode has dephased.

\begin{figure}
	\centering
	\includegraphics[width=0.49\textwidth]{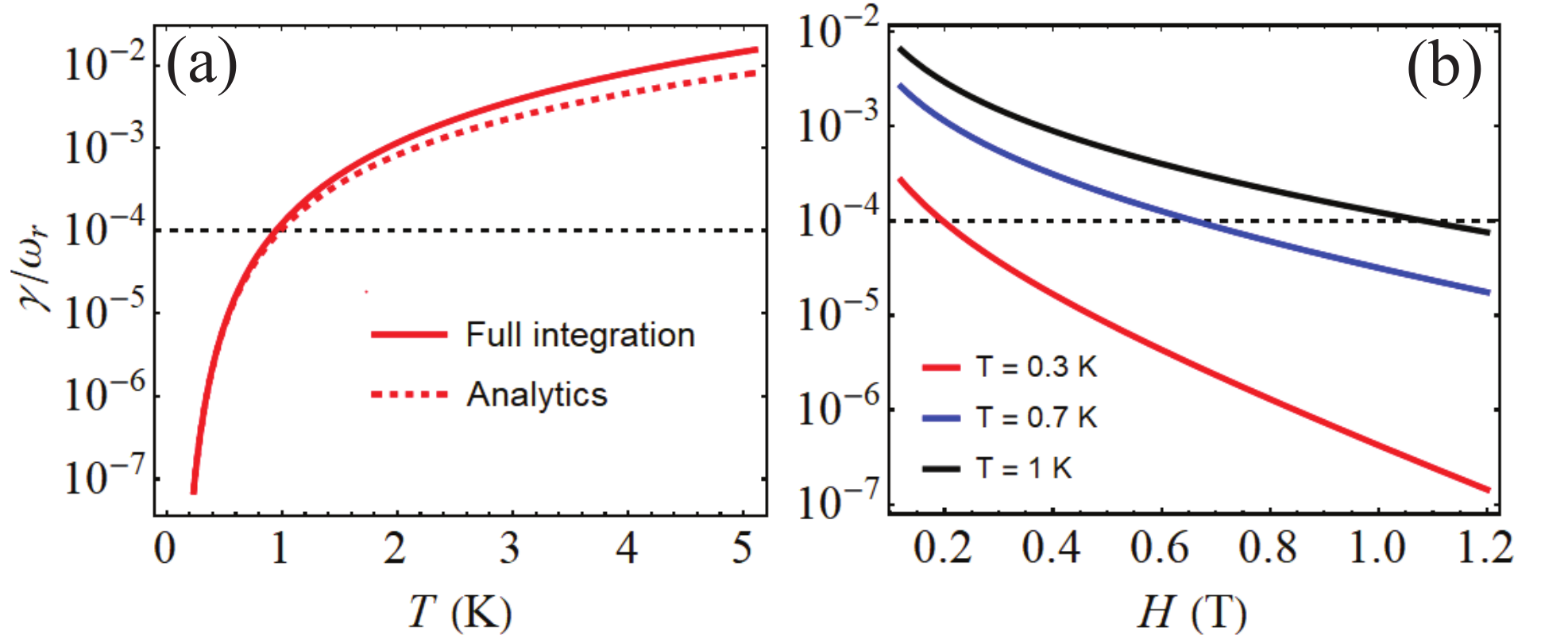}\\
	\caption{(a) Magnon dephasing rate $\gamma$ normalized by the mode frequency $\omega_r$ as a function of temperature. The solid and dashed lines represent the results of full integration \eqref{exchange_integral} and approximate analytical formula \eqref{exchange_analytics}, respectively. $H=1$ T. (b) Normalized magnon dephasing rate as a function of the external magnetic field at temperature $T=0.3$ K (red line), 0.7 K (blue line) and 1 K (black line), respectively. The full integration of \eqref{exchange_integral} is performed to evaluate the dephasing rate. For a comparison, the uniform magnon mode relaxation rate in yttrium iron garnet around $10^{-4}\omega_r$ is plotted as a horizontal dashed line.}\label{fig3}
\end{figure}

Let us now quantify the dephasing rate in typical magnets. The density of states in Eq.~\eqref{exchange_integral} is $D(\omega) = 2\pi V/(2JSd^2)^{3/2} \sqrt{\omega - H}\Theta(\omega-H)$, where the step function $\Theta(\omega-H)=1$ for $\omega>H$ and $0$ when $\omega<H$. The coupling coefficient $g(\omega)$ has a complex wavevector dependence, especially when dipolar interaction is included in the Hamiltonian~\cite{KrivosikPRB2010,AkhiezerBook1968}. The order of $g(\omega)$, however, is $J/N$, where $N$ is the number of spins in the system. When dipolar interaction is considered, it will contribute as $M_s/N$ with $M_s$ being the saturation magnetization, which is typically much smaller than the contribution from exchange ($M_s \ll J$) and is, thus, neglected here. At low temperatures, the Bose-Einstein distribution is approximated by $n_{\mathrm{th}}(n_{\mathrm{th}}+1)\approx \exp(-\omega/k_BT)$. With these approximations, we analytically evaluate the integral in Eq.~\eqref{exchange_integral} and obtain the dephasing rate
\begin{equation}\label{exchange_analytics}
\gamma = \frac{(2\pi k_BT)^2}{(2S)^3 Je^{H/k_BT}}.
\end{equation}

Figure \ref{fig3}(a) plots the dephasing rate showing it to increase with temperature using the parameters of yttrium iron garnet (YIG) with $J=1.59~\mathrm{K}$ and $S=14.2$ \cite{RuckriegelPRB2014}. This trend is well reproduced by both the numerical integration of Eq.~\eqref{exchange_integral} and our analytical formula \eqref{exchange_analytics}. At higher temperatures ($T>2$ K), the analytical result deviates considerably from the full integral [Fig.~\ref{fig3}(a)]. Figure \ref{fig3}(b) shows the decrease of the magnon dephasing with the increase of external field. This is because an external magnetic field enhances the precessional motion of the spins and makes the system more robust against the frequency perturbations caused by coupling to the bath. A typical value for the dephasing rate at $T=1~\mathrm{K}, H=0.1~\mathrm{T}$ evaluated as $\gamma = 8.4 \times 10^{-3} \omega_r$ is much larger than the magnon relaxation rate of $\sim 10^{-4} \omega_r$ in the widely used YIG sphere. Hence, magnon dephasing is expected to play an important role in the evolution and stability of magnonic quantum states, even at low temperatures.


{\it Dephasing via magnon-phonon interaction.---} Magnons can also dephase due to their coupling with the phonon bath, which exists in all materials. Assuming a cubic crystal, the magnetoelastic interaction reads~\cite{KittelRMP1949,DreherPRB2012,RuckriegelPRB2014,KamraSSC2014}
\begin{equation}\label{ham_me}
\hat{\mathcal{H}}_{\mathrm{int}}=\sum_\mathbf{r} \sum_{pq} [b_{pq} \hat{S}_p(\mathbf{r}) \hat{S}_q(\mathbf{r}) + b_{pq}^{'}\partial_p \hat{\mathbf{S}}(\mathbf{r}) \cdot \partial_q \hat{\mathbf{S}}(\mathbf{r})]\hat{\epsilon}_{pq},
\end{equation}
where $p,q=x,y,z$, the strain tensor $\hat{\epsilon}_{pq}$ is defined in terms of the lattice displacements $\hat{\mathbf{u}}$ as $\hat{\epsilon}_{pq} \equiv (\partial_q \hat{u}_p+\partial_p \hat{u}_q)/2$, $b_{pq}$ and $b_{pq}^{'}$ are the magnetoelastic coupling coefficients. As a specific case, we consider pure dephasing of the uniform magnon mode in a one-dimensional (1D) spin chain. By considering the case that atoms significantly vibrate in one dimension ($z$-axis), the only nonvanishing term in Eq. \eqref{ham_me} is $\hat{\mathcal{H}}_{\mathrm{int}}=\sum_i b_{zz}\hat{S}_{iz}^2\hat{\epsilon}_{zz}$. The atom displacement field is quantized as $\hat{u}_z=\sum_k (2\rho \omega_kV)^{-1/2}(\hat{b}_k + \hat{b}_k^\dagger)e^{ikz}$, where $\rho$ and $V$ are respectively the mass density and volume of the magnet, $\hat{b}_k~(\hat{b}_k^\dagger)$ is the annihilation (creation) operator of a phonon state with wavenumber $k$ and frequency $\omega_k$, and $\omega_k=c|k|$ is the dispersion of acoustic phonons with $c$ being the longitudinal phonon speed. By substituting the quantized form of magnons and photons into the interaction Hamiltonian $\hat{\mathcal{H}}_{\mathrm{int}}$, we obtain
\begin{equation}
\hat{\mathcal{H}}_{\mathrm{int}}= \hat{a}^\dagger \hat{a}\sum_k g(\omega_k) ( \hat{b}_k  - \hat{b}_k^\dagger ),
\end{equation}
where $g(\omega_k)=-2iSb_{zz}k/\sqrt{2\rho \omega_k V}$ is the frequency-dependent coupling coefficient. Here we have released the requirement of momentum conservation. This may be caused by impurities, grain boundaries and other inhomogeneities in the system \cite{MichaelJAP2002, Safonov2003}. Finally, we obtain the total Hamiltonian as
\begin{equation}
\hat{\mathcal{H}}=\omega_r \hat{a}^\dagger \hat{a} + \sum_k \omega_k \hat{b}_k^\dagger \hat{b}_k +\hat{\mathcal{H}}_{\mathrm{int}}.
\end{equation}
Following the same methodology as employed in treating magnon-magnon interactions above, we arrive at the same master equation \eqref{master_equation}, but now with the dephasing rate
\begin{equation}
\gamma'=\int_0^\infty d \omega D(\omega) |g(\omega)|^2(2n_\mathrm{th}+1)\delta(\omega).
\end{equation}
In 1D case, the phonon density of states $D(\omega)=2V/(cd^2)$ with $d$ being the lattice constant, we can analytically evaluate the integral as
\begin{equation}
\gamma'=\frac{8(Sb_{zz})^2k_BT}{\rho d^2 c^3}.
\end{equation}
The dephasing rate thus obtained depends linearly on the temperature, which is different from the dephasing caused by magnon-magnon interaction. This may help experiments discriminate between the two dephasing channels considered here. Similar to our considerations above, we interpret this dephasing mechanism as due to the phase broadening caused by a randomly-fluctuating phonon-mediated contribution to the magnon frequency. For YIG~\cite{RuckriegelPRB2014}, $b_{zz}=994~\mathrm{GHz}, c=7209~ \mathrm{m/s}, d=1.2376 ~\mathrm{nm}, \rho = 5172 ~\mathrm{kg/m^3}$, the dephasing rate at $T=1$ K is evaluated as $\gamma'=6.7 \times 10^{-5}\omega_r$ for $H=0.1$ T. This value is smaller than the contribution from exchange interaction, but it is still comparable to the relaxation rate of the uniform magnon mode in millimeter-sized YIG spheres~\cite{HarderPRL2018}, and may dominate the decoherence process as the temperature increases further.

{\it Discussion and conclusion.---} We have studied pure dephasing channels of magnons through magnon-magnon and magnon-phonon interactions, and identified their temperature dependencies. Such a difference is rooted in the distinct properties of three-particle and four-particle interaction in these two channels. These two types of interaction should be sufficiently general to cover a large class of dephasing processes, where the methodology presented here can be readily applied to other situations. For example, a ferromagnet under a longitudinal incoherent driving should also suffer from pure dephasing, and it may resemble the dephasing via magnon-phonon interaction.

A measurement of the magnon dephasing time requires examining the decoherence of magnonic nonclassical states, such as a single-magnon state. However, relaxation, on top of pure dephasing, also contributes to the decoherence of a quantum state \cite{NazarovBook2009}, which indicates that the total decherence time $T_2 < T_2^*, T_1$ with $T_1$ the relaxation time of magnons. We outline a possible method for determining the pure magnon dephasing rate when it is comparable to the relaxation rate. First, we can calibrate the absorption linewidth of a magnetic sphere by the technique of ferromagnetic resonance, through which we deduce the relaxation time $T_1=1/(\alpha \omega_r)$ with $\alpha$ being the Gilbert damping of the system \cite{Gilbert2004}. Then we prepare a quantum state of magnons and detect its decoherence time $T_2$ by entangling the magnons with cavity photons and preforming measurements on the cavity output \cite{YuanArxiv2021,Lachance-QuirionScience2020,SharmaPRB2021,SunPRL2021}. After subtracting the influence of the relaxation, we have the contribution of pure dephasing as
\begin{equation}
\frac{1}{T_2^*} = \frac{1}{T_2} - \frac{1}{T_1}.
\end{equation}
This relation resembles the spin decoherence in nuclear spin resonance~\cite{HornakBook1996} as well as considerations of qubits.

In conclusion, we have shown that the processes conserving magnon number contribute to dephasing of magnonic quantum states, while not affecting their relaxation. By accounting for these processes in the density matrix dynamics, we have demonstrated that they play an important role in extinguishing quantum superpositions of magnonic nonclassical states. Our estimates of the dephasing rate resulting from exchange and spin-phonon interactions show that they exceed the relaxation rate in the low temperature regime, and thus performing quantum operations within the dephasing time of magnons will be critical for information processing. To maintain the magnon coherence for long times, one can keep lowering the temperature and increase the external field. Choosing magnetic materials with strong exchange and weak magnetoelastic coupling will also be beneficial. Further, it would be meaningful to study dephasing in antiferromagnets, and in particular its influence on the entanglement of sublattice magnons, as well as magnon dephasing in ultrafast processes before magnon relaxation prevails.


\begin{acknowledgments}
 H.Y.Y acknowledges the European Union's Horizon 2020 research and innovation programme under Marie Sk{\l}odowska-Curie Grant Agreement SPINCAT No. 101018193. A.K. acknowledges financial support from the Spanish Ministry for Science and Innovation -- AEI Grant CEX2018-000805-M (through the ``Maria de Maeztu'' Programme for Units of Excellence in R\&D). R.A.D. is member of the D-ITP consortium that is funded by the Dutch Ministry of Education, Culture and Science (OCW). R.A.D. has received funding from the European Research Council (ERC) under the European Union's Horizon 2020 research and innovation programme (Grant No. 725509).
\end{acknowledgments}


\bibliography{magnon_dephasing}

\end{document}